\documentclass[letter]{aa}
\usepackage[varg]{txfonts}
\usepackage{graphicx}
\usepackage{amsmath}

\DeclareUnicodeCharacter{2003}{}
\DeclareUnicodeCharacter{2212}{}

\begin{document}

\title{Gravity in the local Universe: Density and velocity fields using CosmicFlows-4}

\author{H.M.~Courtois\thanks{helene.courtois@univ-lyon1.fr}\inst{1}\and 
A.~Dupuy\inst{2}\and
D.~Guinet\inst{1}\and
G.~Baulieu\inst{1}\and
F.~Ruppin\inst{1} \and
P.~Brenas\inst{1}
}

\institute{Universit\'e Claude Bernard Lyon 1, IUF, IP2I Lyon, 69622 Villeurbanne, France\\
\and Korea Institute for Advanced Study, 85, Hoegi-ro, Dongdaemun-gu, Seoul 02455, Republic of Korea
} 

\date{Received A\&A Oct 31, 2022 - AA/2022/45331; Accepted date }

\abstract {This article publicly releases 3D reconstructions of the local Universe  gravitational field below z=0.8 that were computed using the CosmicFlows-4 (CF4) catalog of 56,000 galaxy distances and its subsample of 1,008 supernovae distances. The article also provides measurements of the growth rate of structure using the pairwise correlation of radial peculiar velocities $f \sigma_8 = 0.38(\pm 0.04)$ (ungrouped CF4), $f \sigma_8 = 0.36(\pm 0.05)$ (grouped CF4), $f \sigma_8 = 0.30(\pm 0.06)$ (supernovae), and of the bulk flow in the 3D-reconstructed local Universe $230 \pm 136$ km $s^{−1}$ at 300 h$_{100}^{-1}$ Mpc of distance from the observer. The exploration of 10,000 reconstructions has led to the conclusion that the distances delivered by the CF4 catalog are compatible with a Hubble constant of $H_0=74.5 \pm 0.1 $ (grouped CF4), $H_0=75.0 \pm 0.35 $ (ungrouped CF4), and $H_0=75.5 \pm 0.95 $ (CF4 supernovae subsample).
 }


\titlerunning{Gravity in the Local Universe}
\authorrunning{Courtois et al.} 
\maketitle

%

\section{Introduction}
Peculiar (i.e., gravitational) velocities of galaxies are a robust probe for the search of dark matter on the large scales in the Universe. Their radial component can be computed in a basic way directly from galaxy distances. This method is immensely prone to a variety of Malmquist biases. In order to map the local dark matter distribution and to measure various cosmological parameters, modern cosmologists would rather use a full reconstruction in 3D of peculiar velocities. Such reconstructions are based on the Wiener filter algorithm, as well as forward modeling of the dataset and likelihood. Unfortunately, very few public releases of 3D peculiar velocity reconstructions are available to date, the largest one being the reconstruction from the redshift survey 2MASS by \cite{2011MNRAS.416.2840L}. In this article, about 56,000 galaxy distances and 1,000 supernovae distances from the CosmicFlows-4 (CF4) catalog are used to publicly release 3D reconstructions of the local Universe gravitational field. 

The purpose of producing 3D reconstructions is not solely to create maps and cosmography of the nearby large-scale structures. The grids can also be used to test some cosmological hypothesis such as the general relativity model for gravity via the growth rate of large-scale structures (see, for example, \cite{2012ApJ...751L..30H}, \cite{2019MNRAS.486..440D}) and the homogeneity scale of the Universe via a test of the mean of all gravitational velocities enclosed in a sphere, the bulk flow. Both of these cosmology measurements are very sensitive to the number density of the catalogs and to the robustness and accuracy of the observed distance moduli.

For more than a decade, the measurement of distances of supernovae promises to the cosmologist more accuracy on distance moduli and more recently on bulk flow measurements than the classic galaxy distance relations. Already some literature exists that uses up to a few hundred supernovae distances to compute their peculiar velocities (without reconstruction) and to derive a measurement of the local Universe bulk flow.\ For more details, readers can refer to
\cite{2011JCAP...04..015D}, \cite{2012MNRAS.420..447T}, \cite{2013A&A...560A..90F}, \cite{2020MNRAS.498.2703B}, \cite{2021EPJST.230.2067M}, and \cite{2021arXiv211003487P}, for example.
In this article we comparatively study galaxy and supernovae distances as probes of both the growth rate of structure $ f \sigma_8$ and the bulk flow.

\section{Data}

The fourth release of the CosmicFlows catalog \citep{CF4} provides about 56,000 measurements of galaxy distances and about 1,000 supernovae distance moduli measurements. Such a composite catalog of distances delivers the raw material to compute radial peculiar velocities. Since the first CosmicFlows catalog, our peculiar velocity computational tools have evolved from direct analysis of Malmquist-biased radial peculiar velocities (CF1 1,600 galaxies), to the Wiener filter linear 3D-reconstructed dataset (CF2: 8,000 galaxies), which allowed for some modern cosmography of our local Universe to be built \citep{2013AJ....146...69C}, to a forward modeling iterative procedure for the third, much larger dataset (CF3: 18,000 galaxies). CF3 reached greater distances in the southern terrestrial hemisphere and was used to discover distant features such as the Cold Spot Repeller \citep{2017ApJ...847L...6C} and the Vela Supercluster \citep{2019MNRAS.490L..57C}. The current CF4 dataset is three times larger in terms of the number of galaxies than CF3 and it is doubling its reach in the northern hemisphere.

\section{Growth rate measurement from radial peculiar velocities}

In this article, we use a velocity statistical indicator, noted $\psi_1$, introduced by \cite{1989ApJ...344....1G}. This indicator only depends on radial peculiar velocities, and $\psi_1$ is defined as follows:

\begin{equation}
\psi_1(r) = \frac{ \sum{ \vec{u}_A \cdot \vec{u}_B }}{ \sum{ \left( \hat{\vec{r}}_A \cdot \hat{\vec{r}}_B \right)^2 } } = \frac{ \sum{u_A u_B \cos\theta_{AB}} }{ \sum{\cos^2\theta_{AB}} }
\label{eq:psi1}
.\end{equation}

\cite{2019MNRAS.486..440D}, using the direct radial peculiar velocities of the CF3 dataset, found $f \sigma_8 = 0.43(\pm 0.03)_\mathrm{obs}(\pm 0.11)_\mathrm{cos. var.}$ out to z=0.05.

\begin{figure}[h!]
\begin{center}
\includegraphics[width=0.9\columnwidth,angle=-0]{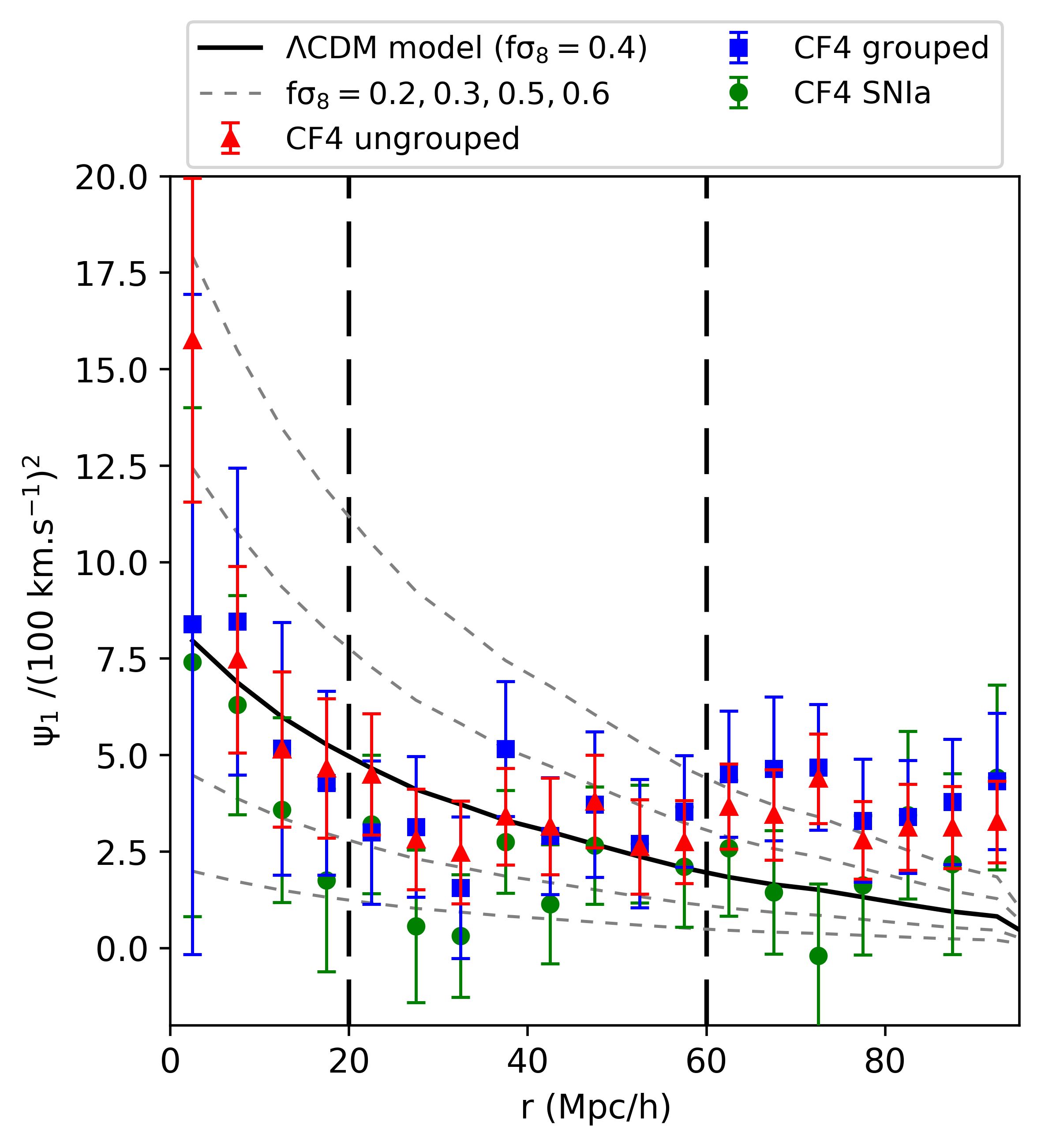}
\caption{Estimator to compute the growth rate of structures using 1,008 supernovae (in green), 36,000 galaxy groups (in blue), and 56,000 individual galaxy distances (in red). Computation was carried out as in \cite{2019MNRAS.486..440D}. The distance $r$ is the pairwise separation bin in h$_{100}^{-1}$ Mpc between radial peculiar velocities in the datasets. Error bars represent the observational errors of the distance moduli (no cosmic variance is considered in this plot). The number density of data per bin of pairwise separation is of crucial importance in order to assess the result. The current supernovae sample is too sparse for separations smaller than 20--30 h$_{100}^{-1}$ Mpc.}
\label{fig:Phi_1}
\end{center}
\end{figure}

Figure~\ref{fig:Phi_1} shows the computation of the estimator $\Psi_1$ using the radial peculiar velocities of the 1,008 supernovae in CF4 (in green), and of the 36,000 galaxy groups in CF4 (in blue) and the full 56,000 individual galaxies (in red). It appears clearly that the small number of supernovae per bin of pairwise separation is too small for pairwise separations smaller than 20--30 h$_{100}^{-1}$ Mpc. The grouped and ungrouped versions of the CF4 catalog provide results that concord for the two samples across the whole range of pairwise separations.

This procedure has provided the following for the supernovae: $f \sigma_8 = 0.28(\pm 0.09)_\mathrm{obs}$ for the pairwise separation range 20-30 h$_{100}^{-1}$ Mpc, $f \sigma_8 = 0.31(\pm 0.06)_\mathrm{obs}$ for the range 20-60 h$_{100}^{-1}$ Mpc, and $f \sigma_8 = 0.30(\pm 0.06)_\mathrm{obs}$ for the range 20-80 h$_{100}^{-1}$ Mpc.  We obtain  $f \sigma_8 = 0.32(\pm 0.07)_\mathrm{obs}$ for the grouped CF4 and $f \sigma_8 = 0.36(\pm 0.06)_\mathrm{obs}$ for the individual galaxies in CF4, with both being for the pairwise separation range 20-60 h$_{100}^{-1}$ Mpc. All errors were computed using the observational errors of the distance moduli. Cosmic variance is much larger and in \cite{2019MNRAS.486..440D} we determined that it would account for about $\pm 0.11$ in the volume of the Universe covered by CosmicFlows catalogs.
This value should be compared to the y axis of Figure 7 in that paper. Here, we recomputed the growth rate with the same methodology applied to CF4 radial peculiar velocities, and error bars represent observational errors only. However, the error bars do include the number density in each bin. The two-point velocity statistic considered here allows for the underlying cosmology from a homogeneous and spherical universe to be computed. As tested and explained in \cite{2019MNRAS.486..440D}, the unique geometry of the CF3 and CF4 catalogs would impact the findings about the underlying cosmology for pairwise separation distances larger than 60 Mpc/h.

\section{3D gravitational velocity field and bulk flow}

In order to compute a 3D reconstruction of the CF4 catalog, we used an iterative forward modeling procedure with a Hamiltonian Monte-Carlo (HMC) algorithm that explores the values of some free parameters  (i.e., $\Omega_m$, bias, scatter component of the nonlinearity in the velocity field solution $\sigma_{NL}$, among others). This procedure is an extension of the procedure used for the CF3 catalog described in \cite{Graziani2019} and briefly hereafter.

To obtain the linear particular velocity field, the following linearized continuity equation was used:\\
\begin{equation}
\frac{\partial\delta}{\partial t} + \frac{1}{a}\boldsymbol{\nabla}_{\chi} .\textbf{v} = 0
,\end{equation}
where the over-density field is 
\begin{equation}
    \delta(\textbf{x},t) = D_+(t)\delta_0(\textbf{x}),
    \label{eq:delta_lin}
\end{equation}
where  $D_+(t)$ is the linear growth factor. Indeed, the linear approximation carried out in this article gives the linear temporal evolution of the over-density field. The continuity equation then becomes the following:
\begin{equation}
    \boldsymbol{\nabla}_{\chi}.\textbf{v} = -\dot{D}\delta_0 = -\frac{d\ln(D)}{d\ln(a)}\dot{a }\delta = -\dot{a}f\delta,
\end{equation}
where $f = \frac{d\ln(D)}{d\ln(a)}$ is the growth rate of large scale structures.
The power spectrum $P(\textbf{k})$ is the function that characterizes the amplitude of the perturbations in Fourier space, as a function of the scale of these disturbances. The velocity power spectrum corresponds to the over-density power spectrum times $\frac{1}{k^2}$.
This means that the peculiar velocity field is less affected by small-scale changes and is more strongly correlated at large scales. \\
The Bayesian model used here is an adaptation of a model developed by \cite{Lavaux2016}. 
Two observations are used for each galaxy in the catalog: the distance modulus $\mu$ and the observed velocity $v_{CMB}=cz_{obs}$.\\
The conditional probability of the distance modulus knowing the parameters is the following:\\
\begin{equation}
    P(\mu|d) = N(5\log(\frac{h_{\mathrm{eff}}d}{10pc}),\sigma_{\mu}^2;\mu).
    \label{eq:p(mu)}
\end{equation}
The $\mu$ probability is a Gaussian centered around $ 5 \log(\frac{h_{\mathrm{eff}}d}{10pc})$ and with standard deviation $\sigma_{\mu}$, which corresponds to the error of the measurement of the distance modulus.The parameter $h_{eff}$ is quite important in the model. It traces zero-point intercalibration problems between different distance methodologies used in the CosmicFlows catalogs. The relationship between the distance modulus and the luminosity distance is valid up to a constant. This constant is the zero point of the law that indicates the distance and which must be calibrated. This calibration is not independent of the value of $H_0$ and its value could influence the reconstruction performed.
The conditional probability of the observed redshift knowing the parameters is given by the following:\\
\begin{equation}
    P(z|\boldsymbol{v},d,\sigma_{\mathrm{NL}},\boldsymbol{r}) = 
    N\Big(\boldsymbol{v}.\textbf{r} , \sigma^2_{ cz}(1+\bar{z})^{-2} + \sigma^2_{NL} ,v^r_{pec}\Big),
    \label{eq:p(z)}
\end{equation}
where $v^r_{pec}$ is the radial peculiar velocity. It is a Gaussian law centered around the reconstructed velocity for each point of the grid, with its standard deviation being a combination between the typical error of the observed redshift $\sigma_{cz}$ and a $\sigma_{NL}$ parameter. The purpose of the latter is to absorb the nonlinearities of the observed velocities.\\
The likelihood function was then constructed as the product of the conditional probabilities in equations \ref{eq:p(mu)} and \ref{eq:p(z)}.
Priors are functions that assign a probability to each state in the parameter space. These functions guide the sampling toward high probability parameter values and, when chosen carefully, accelerate the convergence of the sampling.\\
In the case of the model used here, the parameters for which a prior must be chosen are $h_{\mathrm{eff}}$, $\sigma_{\mathrm{NL}}$, $\{d_i\}$, and $ \boldsymbol{v}$. The particular velocities are known a priori thanks to the power spectrum of the chosen cosmological paradigm (here LCDM from \cite{Planck2015}), which makes it possible to define an a priori over-density field: \\
\begin{equation*}
P(\delta(\boldsymbol{k})|\Lambda CDM) = \prod_{\boldsymbol{k}} N(0,P(\boldsymbol{k})).
\end{equation*}
This prior for the over-density field allowed us to derive the prior for the peculiar velocity field in Fourier space using linear perturbation theory. This realization of the velocity field was then constrained by the observational data using the Wiener filter constrained realizations as described,  for example, in \cite{Doumler2013}. This work provides a brief overview on how to reconstruct the 3D peculiar velocity field.\\
The reconstruction codes we have used are continuously improved by our team and thoroughly tested. The one used for this article already shows limitations: namely in the time needed to compute reconstructions of large datasets such as CF4. Thus we are also testing newer reconstruction methods that will allow us to analyze the upcoming datasets of WALLABY, DESI, and 4HS peculiar velocity surveys, which will be an order of magnitude larger than CF4.

\begin{figure*}[h!] 
\includegraphics[width=1.3\columnwidth,angle=-0]{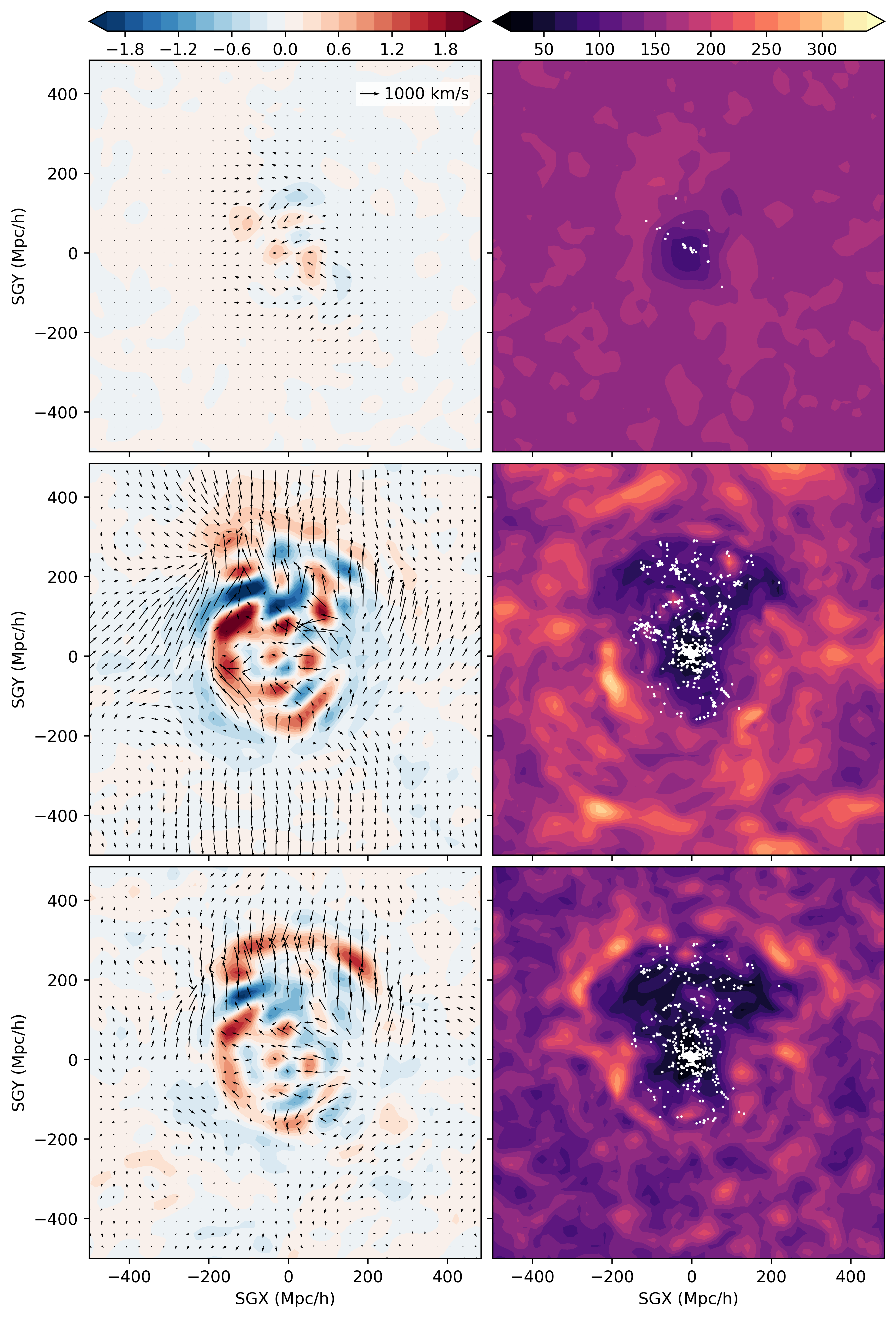}
\caption{Over-density $\delta$ and velocity gravitational fields 3D reconstructions of the CF4 supernovae sample (top row), full galaxy sample (ungrouped and grouped in the middle and bottom row, respectively), projected onto the supergalactic coordinate SGX-SGY plane (Left column). Velocity field standard deviation in km $s^{-1}$ over the 10,000 HMC steps of the respective computations of the left panels (Right column). The white dots are the CF4 data in a slice of $\pm$ 2h$_{100}^{-1}$ Mpc on the SGZ axis.  In the nearby Universe close to the observer, located at the center of the box, all datasets reconstruct the same large-scale structures, with a  robustness in the peculiar velocity field of about 50 km $s^{-1}$. The Sloan great wall appears at SGY 250 h$_{100}^{-1}$ Mpc as a major feature of our local Universe. }
\label{fig:SGX-SGY}
\end{figure*}


Since CF4 is made up of several surveys, an additional step is needed to prevent Malmquist biases and zero-point calibration errors. This prior is an arbitrarily chosen function that models the expected behavior of the distance distribution. We followed \cite{Lavaux2016} by fitting a function with three parameters, $a$, $b$, and $c$, to the distribution of redshifts. Figure \ref{fig:priors} shows that the catalog contains two classes of supernovae and also two classes of galaxy distances: a class of data with high accuracy (labeled type 0) with errors for distance moduli being smaller than 0.5 mag (in blue) and a medium accuracy class of data (labeled type 1) with errors for distance moduli being larger or equal to 0.5 mag (in green).

In this article all computations have been carried out with the $\Lambda CDM$ cosmological model with initial values of $\Omega_m$ = 0.3, $H_0$ = 74.6 km $s^{-1}$ Mpc$^{-1}$ (the nominal value for the CF4 catalog given in \cite{CF4}), and a limitation of $k_{max}$ =0.1 for the smallest phases reconstructed in Fourier space. In order to reach a stable convergence on the parameters, the computation requires approximately 3,000 HMC steps for both the supernovae subsample and the full CF4 dataset. However, we have gone well beyond the convergence and computed about 10,000 HMC steps in order to subsequently derive mean solutions that are less correlated and more statistically robust.
The resulting over-density $\delta$ and gravitational velocity fields were averaged (after removing the burning steps of the HMC). The resulting velocity fields are presented in this article for resolutions of $64^3$ up to $z=0.08$ for the CF4 dataset and its supernovae subsample. This yields about $8 \mathrm{h^{-1}Mpc}$ per voxel. A voxel is the cubic volume in a 3D density grid. Its size depends on the resolution of the grid.
The nonlinear component of the velocity field, which is a free parameter of the HMC exploration, has a mean value of $\sigma_{NL}^{CF4gp}$=170 km $s^{-1}$, $\sigma_{NL}^{CF4}$=300 km $s^{-1}$, and  $\sigma_{NL}^{supernovae}$=250 km $s^{-1}$.


The exploration of the 10,000 reconstructions has led to the conclusion that the distances delivered by the CF4 catalog are compatible with a Hubble constant of $H_0=74.5 \pm 0.1 $ (grouped CF4), $H_0=75.0 \pm 0.35 $ (ungrouped CF4), and $H_0=75.5 \pm 0.95 $ (CF4 supernovae  subsample).

Figure~\ref{fig:SGX-SGY} shows the reconstructed 3D density $\delta_m$ and velocity $\boldsymbol{v_m}$ gravitational fields for the three datasets. 
It is important that the same large-scale structures be found within the first 100 h$_{100}^{-1}$ Mpc around the observer with all datasets. Those are the first published 3D reconstructions of CF4. The largest structure known in the local Universe, the Sloan great wall, is clearly seen at SGY of 250 h$_{100}^{-1}$ Mpc and extends over more than 400 h$_{100}^{-1}$ Mpc along the SGX coordinates.
One of the improvements compared to previous CosmicFlows reconstructions is that now we also deliver the standard deviation of the peculiar velocity field over the HMC forward modeling process. In fig.~\ref{fig:SGX-SGY}, the right panels show this standard deviation in km s$^{-1}$. Nearby, in a radius of about 200 h$_{100}^{-1}$ Mpc around the observer located at (SGX=0, SGY=0), the reconstruction we publish is robust at the level of 50 to 80 km $s^{-1}$.



From the 3D velocity field, we computed a basic estimation of the bulk flow as the mean of the gravitational velocities, which are at rest in the cosmic microwave background, inside a sphere of radius $R$. The errors of the bulk flow were computed from the standard deviation of the peculiar velocity field obtained through the variations of each of the realizations during the HMC forward modeling process.

\begin{figure}[h!]
\begin{center}
\includegraphics[width=0.95\columnwidth,angle=-0]{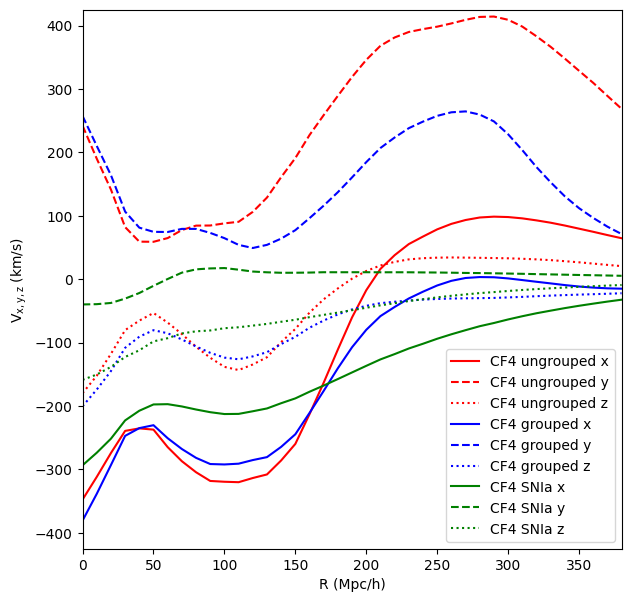}\\
\includegraphics[width=0.95\columnwidth,angle=-0]{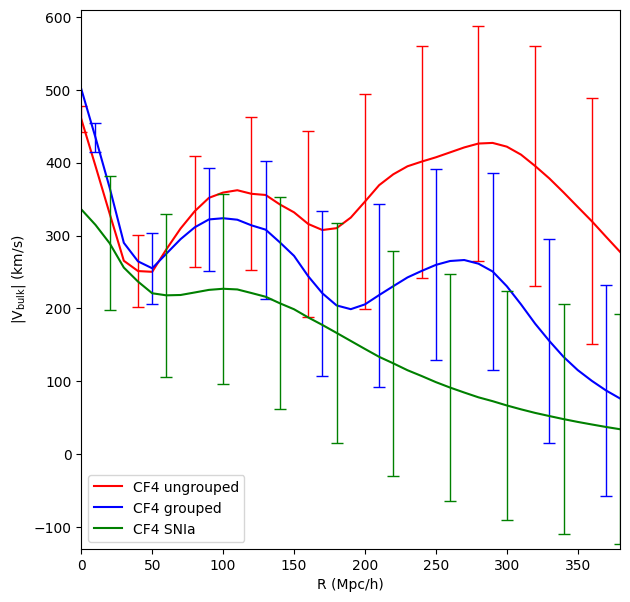}
\caption{Bulk flow components along the supergalactic X, Y, and Z axis (top panel) and its norm (bottom panel), as a function of the radius $R$ of a sphere centered at the observer location. The velocities are in km $s^{-1}$ at rest in the cosmic microwave background. The error bars for the 3D reconstruction were computed from the standard deviations of 10,000 HMC realizations. The reader is reminded that the supernovae dataset only extends out to 100 h$_{100}^{-1}$ Mpc and it is too sparse to fully recover the velocity field.}
\label{bulkNorm}
\end{center}
\end{figure}

Figure~\ref{bulkNorm} shows the amplitude of the bulk flow (bottom panel) and of its three components along the supergalactic X, Y, and Z axis (top panel), as a function of the radius of the sphere centered at the observer location.  The error bars for the 3D reconstruction were computed from the standard deviations on the 10,000 HMC realizations. At large radii, where the prior model of an homogeneous universe dominates over the observational data, the bulk flow vanishes. Both the supernovae and the full CF4 sample concord and detect the large nearby attractors located within [100-150]~h$_{100}^{-1}$ Mpc from the observer: Centaurus, Great Attractor, and Coma. Inside 150 h$_{100}^{-1}$ Mpc, the CF4 bulk flow is $(272\pm 105)$ km $s^{-1}$, which is in complete accordance with our previous measurement of $(239 \pm 38)$ km $s^{-1}$ using the CF2 dataset \citep{2015MNRAS.449.4494H}. Using 1,008 supernovae, we find a bulk flow of $227 \pm 131$ km $s^{-1}$ at 100h$_{100}^{-1}$ Mpc (see Table~\ref{Table_bulk}), which is in agreement with \cite{2012MNRAS.420..447T} using 245 supernovae peculiar velocities who measured a bulk flow of $249 \pm 76$ km $s^{-1}$ out to 110 h$_{100}^{-1}$ Mpc. 

Further out from the observer, the impact of the Sloan great wall is clearly seen in fig.~\ref{bulkNorm} with a strong pull along the SGY direction within the range of distances, [250-300]~h$_{100}^{-1}$ Mpc, contributing to a bulk flow of 230 km $s^{-1}$  at the largest distance tested, 300 h$_{100}^{-1}$ Mpc. This value is in strong disagreement with
\cite{2010ApJ...712L..81K}, who found -- using X-ray emission of clusters of galaxies -- a bulk flow of about $1,000 \pm 300$ km $s^{-1}$ that is constant across the range [175-525] h$_{100}^{-1}$ Mpc.

\begin{table}
\begin{tabular}{lrrrr}
\hline
Radius  & $V_{Bulk}$ Norm  & Vx  & Vy  & Vz \\
h$_{100}^{-1}$ Mpc & km $s^{-1}$ & km $s^{-1}$ & km $s^{-1}$ & km $s^{-1}$ \\
\hline
Ungrouped CF4 \\
\hline
  50 &   250 $\pm$ 57 &   -237 &   59 &   -53\\
  100&   359 $\pm$ 89 &   -319 &   88 &   -138\\
  150&   332 $\pm$ 123 &   -260 &   191 &   -78\\
  200&   347 $\pm$ 148 &   -17 &   346 &   13\\
  250&   407 $\pm$ 160 &   79 &   398 &   34\\
  300&   422 $\pm$ 162 &   98 &   409 &   33\\
\hline
Grouped CF4\\
\hline
  50 &   255 $\pm$ 49 &   -230 &   75 &   -80\\
  100&   324 $\pm$ 78 &   -292 &   65 &   -124\\
  150&   272 $\pm$ 105 &   -245 &   77 &   -91\\
  200&   205 $\pm$ 123 &   -80 &   185 &   -42\\
  250&   260 $\pm$ 131 &   -10 &   258 &   -31\\
  300&   230 $\pm$ 136 &   1 &   229 &   -29\\
\hline
supernovae in CF4\\
\hline
  50&    221 $\pm$ 107 &   -197 &   -10 &   -98\\
  100&   227 $\pm$ 131 &   -213 &   18 &   -77\\
  150&   199 $\pm$ 148 &   -188 &   10 &   -64\\
  200&   144 $\pm$ 153 &   -137 &   11 &   -45\\
  250&    99 $\pm$ 156 &    -94 &   11 &  -29 \\
  300&    67 $\pm$ 157 &    -63 &    9 &  -19\\
  \hline
\end{tabular}
\caption{Bulk flow measurements in spheres of radius R centered at the observer location. Values are in the cosmic microwave background rest frame.}
\label{Table_bulk}
\end{table}

\section{Conclusions}

This analysis of the CF4 catalog provides an estimated growth rate of structures of $f \sigma_8 = 0.36 (\pm 0.05)_\mathrm{obs}$, while the similar previous measurement with CF3 was $ f \sigma_8 = 0.43 (\pm 0.03)_\mathrm{obs}(\pm 0.11)_\mathrm{cos. var.}$.  At the largest distance of the observational
data, we obtain a bulk flow measurement of $230 \pm 136$ km $s^{-1}$ at a distance of 300 h$_{100}^{-1}$ Mpc, which is pointing in the direction of the Sloan great wall.
 Another conclusion is that, although the current supernovae sample is very small, it is dense enough for an acceptable local $f \sigma_8$ measurement. However, the supernovae dataset only extends out to 100 h$_{100}^{-1}$ Mpc and is too sparse to fully recover the velocity field and, as a consequence, it cannot be used as a probe of the bulk flow yet. 

From the tight variation around $\Lambda CDM$ values of the free parameters explored by the convergence of the HMC forward modeling and in light of the values obtained for two  cosmology-related parameters -- namely the growth rate and bulk flow -- we conclude that this analysis of the local Universe using peculiar velocities is compatible with a $\Lambda CDM $ cosmology. 

The density and velocity gravitational fields' 3D reconstructions on grids will be publicly available at  \url{https://projets.ip2i.in2p3.fr//cosmicflows/}. We also go one step further, compared to our previous CosmicFlows analysis, by delivering the accompanying 3D standard deviation grids of the reconstructed gravitational fields in order for future users to better estimate  their analysis robustness.

\begin{acknowledgements}
HC is grateful to the Institut Universitaire de France for its huge support which enabled this research. HC, DG and FR acknowledge support from the CNES. AD is supported by a KIAS Individual Grant (PG087201) at Korea Institute for Advanced Study.

\end{acknowledgements}

\bibliography{CF4_gravity}{}
\bibliographystyle{aa}

\begin{appendix}
\section{Appendix A}

\begin{figure} [h!]
\includegraphics[width=0.95\columnwidth,angle=-0]{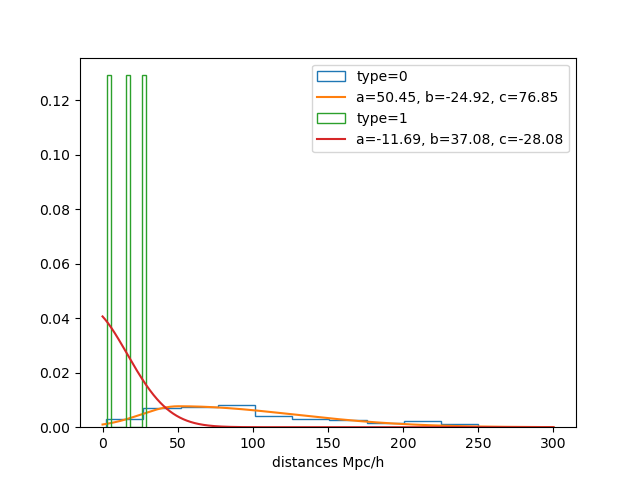}\\
\includegraphics[width=0.95\columnwidth,angle=-0]{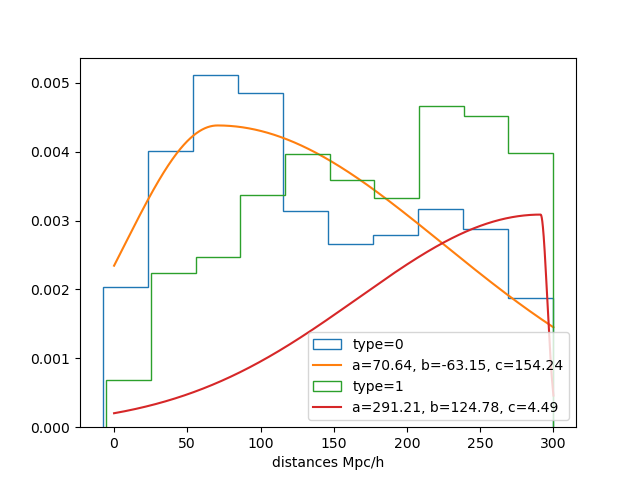}\\
\includegraphics[width=0.95\columnwidth,angle=-0]{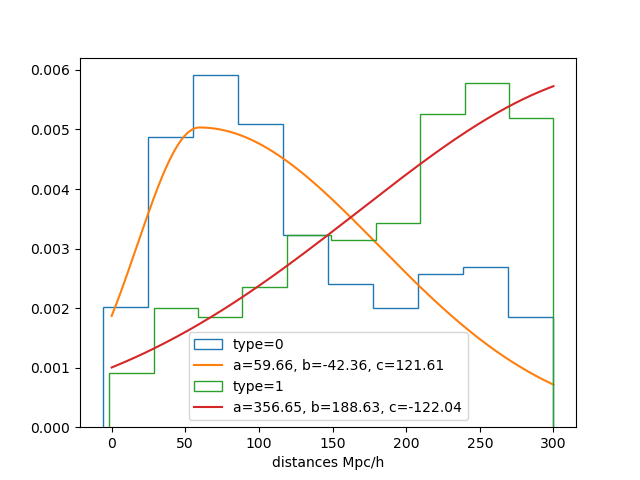}
\caption{Gaussian fitting of the normalized initial distance distribution of the CF4 supernovae  sample (top panel), ungrouped CF4 catalog (middle panel), and grouped CF4 catalog (bottom panel). The priors were fit according to the precision of the distance moduli: less than 0.5 mag errors were modeled as class 0, and $\ge 0.5 mag$ errors are in class 1.}
\label{fig:priors}
\end{figure}
\end{appendix}

\end{document}